\documentclass[nofootinbib,prd,showpacs,preprintnumbers,amsmath,amsfonts,amssymb]{revtex4}

\usepackage[letterpaper,breaklinks]{hyperref}
\usepackage[latin1]{inputenc}
\usepackage{bm}
\usepackage{slashed}

\usepackage[T1]{fontenc}
\usepackage{lmodern} 

\usepackage{graphicx}

\DeclareMathOperator{\tr}{tr}
\renewcommand{\=}{\mathrel{\phantom{=}}}
\newcommand{\s}[1]{\slashed{#1}}
\newcommand{\intt}[2][\,]{\int\!\!\diff[#1]{#2}}
\newcommand{\+}{\negthinspace+\negthinspace}
\newcommand{\mi}{\negthinspace-\negthinspace}
\newcommand{\del}{\partial}

\newcommand{\diff}[2][]{\mathrm{d}^{#1}#2}
\newcommand{\pint}[2][4]{\int\!\!\frac{\diff[#1]{#2}}{(2\pi)^#1}}

\newcommand{\arXiv}[1]{\href{http://arxiv.org/abs/#1}{\texttt{ #1}}}
\newcommand{\narXiv}[2]{\href{http://arxiv.org/abs/#1}{\texttt{arXiv:#1 [#2]}}}

\newcommand{\emPl}[1][]{{M_\text{Pl}^{#1}}}
\newcommand{\kaS}[1][]{{\ka_{\star}^{#1}}}
\newcommand{\emS}[1][]{{M_{\star}^{#1}}}

\renewcommand{\L}{\mathcal{L}}
\renewcommand{\O}{\mathcal{O}}

\newcommand{\al}{\alpha}
\newcommand{\be}{\beta}
\newcommand{\ga}{\gamma}
\newcommand{\de}{\delta}

\newcommand{\ka}{\kappa}

\newcommand{\La}{\Lambda}

\begin{document}
\title{No Lee-Wick Fields out of Gravity}
\date{March 23, 2009}
\author{Andreas Rodigast}
\email{rodigast@physik.hu-berlin.de}
\author{Theodor Schuster}
\email{theodor.schuster@physik.hu-berlin.de}
\affiliation{Humboldt-Universit\"at zu Berlin\\ Institut f\"ur Physik\\ Newtonstra\ss e 15, D-12489 Berlin, Germany}
\preprint{HU-EP-09/13}
\begin{abstract}
We investigate the gravitational one-loop divergences of the standard model in large extra dimensions, with gravitons propagating in the $(4+\delta)$-di\-men\-sional bulk and gauge fields as well as scalar and fermionic multiplets confined to a three-brane. To determine the divergences we establish a cut-off regularization which allows us to extract gauge-in\-va\-ri\-ant counterterms. In contrast to the claim of a recent paper \cite[\href{http://arxiv.org/abs/0807.0132}{\texttt{arXiv:0807.0132}}]{wu2008tsl}, we show that the fermionic and scalar higher derivative counterterms do not coincide with the higher derivative terms in the Lee-Wick standard model. We argue that even if the exact Lee-Wick higher derivative terms were found, as in the case of the pure gauge sector, this would not allow to conclude the existence of the massive ghost fields corresponding to these higher derivative terms in the Lee-Wick standard model.
\end{abstract}
\maketitle

\section{Introduction}
A well-known issue of the standard model is the hierarchy puzzle. The 
mass of the Higgs boson acquires quadratically divergent radiative 
corrections. In order to keep the Higgs mass small compared to the 
Planck scale $M_{\text{Pl}}\sim 10^{19}\text{GeV}$ a delicate 
cancellation has to happen, requiring an extreme fine-tuning.
To solve this problem Grinstein, O'Connell, and Wise 
\cite{grinstein2008lws} suggested a higher derivative generalization of 
the standard model. Their proposition is based on the ideas of Lee and 
Wick \cite{lee1969nma,lee1970ftq}, who studied the consequences of the 
assumption that the modification of the photon propagator in the 
Pauli-Villars regularization \cite{pauli1949irr} of quantum 
electrodynamics, corresponds to a physical degree of freedom. The 
modification of the photon propagator and thus the additional massive 
vector field correspond to a higher derivative term being added to the 
Lagrangian. Exploiting the improved UV behavior Lee and Wick were able 
to construct a finite theory of quantum electrodynamics. Grinstein et 
al.\ extended the standard model to include special dimension-six 
higher derivative terms for each particle. These so called Lee-Wick 
terms have the special property of allowing for an equivalent 
formulation of the theory containing additional massive fields but only 
operators of dimension four or less. This property is crucial for the 
higher derivative theory fulfilling the constraints of perturbative 
unitarity \cite{Grinstein:2007iz}. The new particles are ghosts because 
their kinetic terms have the wrong sign. This indicates an instability 
on the classical level and results in problems with the unitarity for 
the quantum theory. However, these problems appear to be solvable and have been 
extensively discussed in the literature, for example in 
\cite{lee1969nma,lee1970ftq,cutkosky1969nas,vantonder2008uli}.

The Lee-Wick terms used by Grinstein et  al.\ are given by
\begin{equation}\label{eq:LWSM}
\begin{aligned}
 &\tfrac{1}{\vphantom{\hat{M}}M_A^2}\tr\{(D^\mu F_{\mu\nu})^2\} 
&&\text{for gauge fields,}\\
 &\tfrac{1}{\vphantom{\hat{M}}M_{\phi}^2}(D^2\phi)^\dagger(D^2\phi)&&\text{for scalars 
(Higgs), and}\\
&\tfrac{i}{\vphantom{\hat{M}}M_\psi^2}\overline{\psi}\s{D}^3\psi&&\text{for 
fermions}.
\end{aligned}
\end{equation}
This extension, known as the Lee-Wick standard model, is free of 
quadratic divergences and is therefore one possible solution to the 
hierarchy puzzle. Several recent papers investigated the properties of 
the Lee-Wick standard model, e.\,g.\ \cite{LWpheno}.

Shortly after its proposition Wu and Zhong \cite{Wu:2007lw} pointed out a possible 
connection between the Lee-Wick standard model and one-loop 
counterterms in the nonrenormalizable Einstein-Maxwell theory.

It is a well-known fact that quantized general relativity is a 
nonrenormalizable theory 
\cite{thooftVeltman,deser1974nqd,deser1974nqe,deser1974ney}. However, 
nonrenormalizable theories can be renormalized at each loop order 
by including counterterms of higher dimension
and reliable predictions can be made, if they are treated in the general 
enough framework of effective field theories 
\cite{lrr-2004-5,burgess2007ief}. As has been shown in 
\cite{deser1974ney,ebert2008agc,ebert2008gcr}, the only gauge field 
dimension-six counterterm necessary to renormalize 
Einstein-Yang-Mills\footnote{This result also applies to the Abelian 
case.} theory at one-loop order, is the Lee-Wick term \eqref{eq:LWSM} 
for gauge fields. In the case of Einstein-Maxwell theory the same 
interesting observation led Wu and Zhong \cite{Wu:2007lw} to the 
conclusion that gravity provides a mechanism for the emergence of the 
Lee-Wick partner of the gauge field.

The search for a solution of the hierarchy puzzle has also led to models 
modifying gravity itself, e.\,g.\  
\cite{ArkaniHamed:1998rs,Antoniadis:1998ig,Randall:1999ee,Dvali:2000hr,Dvali:2007hz}, 
resulting in a lowering of the characteristic scale of quantum gravity 
from the Planck mass $\emPl$ to an energy $\emS$ which could be as low 
as some TeV without conflicting with experiments.
The most popular of these, the large extra dimension models, in which 
gravitons propagate in the entire space-time bulk whereas the standard 
model matter is confined to a four-di\-men\-sional submanifold 
(three-brane), are of special interest for the gravitational induction 
of Lee-Wick terms. In these models gravity exhibits a much richer 
particle spectrum, i.\,e.\  massive Kaluza-Klein excitations of the graviton and 
additional scalar and vector particles. The more complex structures in 
higher dimensional gravity are expected to be reflected in the gravity 
induced counterterms, including the higher derivative dimension-six 
operators. 

Recently Wu and Zhong \cite{wu2008tsl} claimed that a large extra 
dimension model
provides a mechanism for the emergence of Lee-Wick partners, with 
masses in the TeV scale, for all particles. They base their arguments on the 
higher derivative counterterms that appear in the one-loop 
renormalization of this theory, which according to their results are 
given by the Lee-Wick terms \eqref{eq:LWSM}. However, they only 
calculated two-point functions which, as we will show explicitly, alone 
do not determine the higher derivative counterterms. Consequently, the 
natural question whether or not the fermionic and scalar dimension-six 
counterterms of Einstein-Yang-Mills theory are also given by the 
corresponding Lee-Wick terms \eqref{eq:LWSM}, is still open and will be answered in the following.
\section{Formalism}
We consider a large extra-di\-men\-sional space-time scenario of the form proposed by Arkani-Hamed, Dimopoulos, and Dvali \cite{ArkaniHamed:1998rs}. 
The bulk space-time is the $D$-dimensional manifold $\mathcal{M}=\mathbb{R}^{1,3}\times T^{\de}$, where $T^{\de}$ is a ${\de}$-dimensional torus with a uniform radius $R$ and $\operatorname{dim}\mathcal{M}=D=4+\de$. The graviton moves freely in the bulk whereas the matter and gauge fields are confined on a three-brane, representing the four-di\-men\-sional space-time of the standard model.

In the following upper (lower) case Latin letters are used for $D$-dimensional
($\de$-dimensional compactified) indices and Greek letters for four-di\-men\-sional indices. 
We further decompose the $D$-dimensional coordinates as $X^M=(x^\mu,z^i)$.

The bulk metric is expanded around flat $D$-dimensional Minkowski space-time
\begin{equation}
 G_{M N}(X)=\eta_{M N}+\kaS h_{M N}(X)\,,
\end{equation}
 with $\eta_{MN}=\operatorname{diag}(+,-,\dots,-)$ and the  graviton field $h_{M N}$. 
Here $\kaS$ denotes the gravitational coupling constant in $D$ dimensions which is related by $\kaS[2]=32\pi/\emS[D-2]$ to
the corresponding higher dimensional Planck mass $\emS$. 

The dynamics of gravity are are governed by the Einstein-Hilbert action in the bulk:
\begin{equation}
 S_\text{bulk}=\int\diff[D]{X} \left[\frac{2}{\kaS[2]} \sqrt{\left| G\right|} \,\mathcal{R} - \mathcal{F}_N \mathcal{F}^N + \L_\text{ghosts}\right] . 
\end{equation}
Here $\mathcal{F}_N$ denotes the gauge fixing term and $\L_\text{ghosts}$ the corresponding ghost Lagrangian. A convenient choice is the de~Donder gauge
\begin{equation}
 \mathcal{F}_N=\del^M\left(h_{M N}-\tfrac{1}{2}\eta_{M N}h\right) \qquad\text{with}\qquad	h=h_{MN}\eta^{MN},
\end{equation}
leading to a particularly simple momentum dependence of the graviton propagator and allowing for a direct comparison of our results with those of \cite{wu2008tsl}.
Note that the gravitational Faddeev-Popov ghosts solely couple to gravitons and will therefore play no role in the considered one-loop calculations. Hence, there is no need to specify $\L_\text{ghost}$.

The bulk fields are expanded in eigenmodes of the compactified $\de$ torus (Kaluza-Klein modes)
\begin{equation}
 h_{MN}(x,z) = V_{\delta}^{-1/2} \sum_{\vec{n}\in\mathbb{Z}^\delta} h_{MN}^{(\vec{n})}(x) e^{i\frac{\vec{n}\cdot\vec{z}}{R}} ,
\end{equation}
where $V_{\delta}=(2\pi R)^{\delta}$ is the volume of the compactified torus.

The quadratic part of the graviton action expressed in terms of the Kaluza-Klein modes is
\begin{equation}
\begin{aligned}
 S^\text{quad.}_\text{graviton}=\intt[4]{x}\,\frac{1}{2}\sum_{\vec{n}}\Bigl( \del_\al h^{(\vec{n})}_{M N} &\left(\eta^{M R}\eta^{N S}-
								\tfrac{1}{2}\eta^{M N}\eta^{R S}\right) \del^\al h^{(-\vec{n})}_{R S} - \\
        -m^2_{\vec{n}} h^{(\vec{n})}_{M N} &\left(\eta^{M R}\eta^{N S} -\tfrac{1}{2}\eta^{M N}\eta^{R S}\right)h^{(-\vec{n})}_{R S} \Bigr)\, ,
\end{aligned}
\end{equation}
with the Kaluza-Klein masses $m^2_{\vec{n}}=n^2/R^2$.

To describe matter fields living in four dimensions, we introduce a three-brane, whose position is parameterized by
\begin{equation}
 Y^N(x^\mu)=(y^\mu(x)\!=\!x^\mu,0) \,,
\end{equation}
using a static gauge \cite{Sundrum:1998sj}. Its fluctuations in the compactified directions can be ignored in our considerations, because $\O(\ka^2)$ effects of brane fluctuations yield only seagull graphs \cite{ebert2008gcr}, which do not contribute to the renormalization of the higher derivative operators.

Since brane fluctuations are neglected the induced metric on the brane is simply given by
\begin{align}
 g_{\mu\nu}(x)&= \frac{\del Y^M}{\del x^\mu}\frac{\del Y^N}{\del x^\nu}G_{M N}(Y(x)) \notag \\
              &= G_{\mu\nu}(x,0) \label{eq:indmetric} \\
	      &= \eta_{\mu \nu} + \ka \sum_{\vec{n}\in\mathbb{Z}^\delta} h_{\mu \nu}^{(\vec{n})}(x) \notag\,,
\end{align}
now using the four-dimensional gravitational coupling $\ka= \kaS / \sqrt{V_{\de}}$ observed by particles on the brane. 
The usual definition of the four-dimensional Planck mass $\emPl[2]=32\pi/\ka^2$ yields the relation 
\begin{equation}
 \emS[D-2](2\pi R)^\de =\emPl[2] \,,
\end{equation}
showing that the Planck mass $\emPl\sim 10^{19}\text{GeV}$ can originate from a much smaller gravitational scale $\emS$ in the bulk.

The matter fields and their interactions are described by the brane action
\begin{equation}
 S_\text{brane}=\int\diff[4]{x} \left[ -\tau\sqrt{- g} +\L_\text{YM} +\L_f +\L_s\right] 
\end{equation}
with the constant brane tension $\tau$, which has no one-loop effects on the higher derivative terms, and the Lagrange densities for gauge bosons, fermions, and scalars, which will be specified in the respective sections. 

As the brane Lagrangian, beside the matter fields, only depends on the induced metric \eqref{eq:indmetric}, solely the components $h_{\mu \nu}$ parallel to the brane couple to the matter fields. Thus, we only need the propagator for the components $h_{\mu \nu}$
\begin{equation}\label{eq:grav_prop}
 h_{\al\be}^{(\vec{n})}\;\raisebox{-2.5ex}{\includegraphics{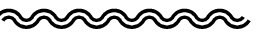}}\;
 h_{\ga\de}^{(\vec{n}')} \: = \: \dfrac{i\de_{\vec{n},-\vec{n}'}\frac{1}{2}\left(\eta_{\al\ga}\eta_{\be\de}+\eta_{\al\de}\eta_{\be\ga} -\tfrac{2}{D-2}\eta_{\al\be}\eta_{\ga\de}\right)}{p^2-m^2_{\vec{n}}}
\end{equation}
in our calculations. Note that in the case of a pure four-di\-men\-sional setup without extra dimensions only the zero mode survives and \eqref{eq:grav_prop} becomes the propagator of a massless spin-2 particle.

\section{Gauge Fields}
The corrections to the dimension-six operators of the vector boson of a non-Abelian gauge theory was already discussed in \cite{ebert2008agc} for $4+0$ dimensions and in \cite{ebert2008gcr} for $d+\de$ dimensions by one of the authors.
For completeness we will quote the previous result. For this consideration the Yang-Mills part of the brane Lagrangian
\begin{equation}
\L_{\text{YM}}=-\tfrac{1}{2}\sqrt{-g}\,g^{\mu\rho}g^{\nu\sigma}\tr\{F_{\mu\nu}F_{\rho\sigma}\} 
\end{equation}
is needed.
In the $(4+\de)$-dimensional case considered in this article, the only gravity induced gauge field counterterm is\footnote{Note that without extra-dimensions, $\de=0$, the power-like divergence becomes logarithmic as\\ $\left(\Lambda^{\de}-\mu^{\de}\right)/\de\rightarrow 1/2 \log \left(\Lambda^2/\mu^2\right)$.}
\begin{equation}\label{eq:gauge_ct}
\L^\text{c.t.}_\text{YM}=
	\frac{1}{(4\pi)^{1+\de/2}\Gamma\left(\frac{\de}{2}+1\right)}\frac{8}{3\de}\frac{\Lambda^{\de}-\mu^{\de}}{\emS[\delta+2]} 
		\tr\left\{ D^\mu F_{\mu\rho} D_\nu F^{\nu\rho} \right\}\,,
\end{equation}
i.\,e.\  there are neither contributions to the dimension-four operator $\tr\{ F_{\mu\nu}F^{\mu\nu}\}$ nor to the further dimension-six operator $\tr\{ F_{\al}^{\phantom{\al}\be} F_{\be}^{\phantom{\be}\ga} F_{\ga}^{\phantom{\ga}\al} \}$.

The absence of gravitational contributions to the dimension-six operator $\tr\{ F^3 \}$ is crucial for the existence of an equivalent formulation \`{a} la Lee-Wick with an additional field and only operators of dimension four or less \cite{grinstein2008lws}.

It should be pointed out, that the values of the dimension-six counterterms depend on the gauge choice in the graviton sector, due to the nonvanishing mass dimension of their coefficients, \cite{Antoniadis:1985ub,Contino:2001nj}. So in a different gauge the $\tr\{ F^3\}$ operator may appear nevertheless. However, all observables, like $S$-matrix elements,  will be independent of the gauge.

Also note that the calculation of the two-point function of an Abelian vector field, as done in \cite{Wu:2007lw} and \cite{wu2008tsl}, is not sufficient to claim the absence of the $\tr\{ F^3\}$ operator in the non-Abelian case. It only manifests itself in three- (and higher) point functions and vanishes identically for Abelian fields. 
\section{Fermions}
To analyze the higher derivative structure of the Einstein-Dirac system it is sufficient to consider a massless fermion on the three-brane. Its Lagrangian is given by
\begin{equation}\label{eq:L_f}
\L_f=\sqrt{-g}\;\overline{\psi}i\s{\mathcal{D}}\psi \,,
\end{equation}
with $\s{\mathcal{D}}=\gamma^a e^{\mu}_{a} \mathcal{D}_{\mu}$, Dirac matrix $\ga^a$, inverse vierbein $e^{\mu}_{a}$, and covariant derivative
\begin{equation}
 \mathcal{D}_\mu=\del_\mu-i\Omega_\mu-igA_\mu \,.
\end{equation}
Here $\Omega_\mu=\tfrac{1}{2}S_{ab}\omega_\mu^{ab}$ is the spin connection with $S_{ab}=\tfrac{i}{4}[\ga_a,\ga_b]$.

Starting from \eqref{eq:L_f} we proceed by expanding the metric dependent quantities around flat space.	The expansion of the Lagrangian to the needed order is given by
\begin{multline}
 \L_f =\overline{\psi}i\s{D}\psi+i\tfrac{\kappa}{2}\,\overline{\psi}[h\s{D}-\gamma^a h_a^\mu D_\mu+\tfrac{1}{2}\partial_b\s{h}_a\gamma^{ab}]\psi 
	+i\tfrac{\kappa^2}{8}\,\overline{\psi}[(h^2\mi 2h^{\alpha\beta}h_{\alpha\beta})\s{D}+(3\gamma^a h^\mu_\rho h^\rho_a-2h\gamma^a h^\mu_a) D_\mu +\\
	+h^{\nu}_b(\partial_a \s{h}_{\nu}\mi\partial_\nu \s{h}_a\+\tfrac{1}{2}\s{\partial}  h_{\nu a})\gamma^{ab} 
	+h\partial_b \s{h}_a \gamma^{ab}-\gamma^a h_a^\mu \partial_c h_{\mu b}\gamma^{bc}]\psi +\O(\ka^3) 
\end{multline}
with $h=h^\mu_\mu$, $\ga^{a b}=\ga^{[a}\ga^{b]}$, and $D_\mu=\del_\mu-igA_\mu$.
It is used to derive the Feynman rules for the fermion propagator and graviton--fermion interactions. For their explicit form see \cite{schuster2008dpl}.
\begin{figure}
\includegraphics{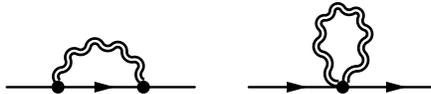}\vspace{-1.2cm}
\caption{One-loop diagrams for the proper fermion two-point function.}\label{fig:ferm_2pt}
\end{figure}

Let us begin with the calculation of the gravitational contributions to the proper fermion two-point function. The relevant one-loop graphs are shown in figure~\ref{fig:ferm_2pt}.
The leading divergence of the two diagrams contributes to the field-strength renormalization of the fermion, and the next-to-leading divergences will renormalize higher derivative operators of mass dimension six as e.\,g.\  the Lee-Wick operator \eqref{eq:LWSM}. For the charged fermion only four of these operators are linearly independent \cite{schuster2008dpl}. As a basis for the counterterms we choose 
\begin{equation}\label{eq:ferm_op}
i\overline{\psi}\s{D}\s{D}\s{D}\psi\,, \qquad i\overline{\psi}\s{D}D^2\psi\,, \qquad i\overline{\psi}D^2\s{D}\psi\,, \quad \text{and}\quad i\overline{\psi}D_\mu\s{D}D^\mu\psi \,.
\end{equation}
The first operator is of the Lee-Wick form and represents new fermionic ghost degrees of freedom. If only this operator is renormalized, the fermionic sector is consistent  with being part of a Lee-Wick field theory, as it is the case for the gauge field. Similarly to the gauge sector all other higher derivative terms have to vanish in order to allow for an equivalent formulation with only operators of dimension four or less \cite{grinstein2008lws}.

To regularize the UV-divergences of the diagrams a cut-off regularization is applied, which involves a cut-off of the $(4+\delta)$-dimensional momentum and a particular parameterization of the loop integrals.
As one can already see in simple examples, in cut-off regularization the regularized value of the nonleading power-like divergences depends on the parameterization of the loop momentum. This ambiguity has to be eliminated to extract the structure of higher derivative counterterms in the extra-di\-men\-sional scenario\footnote{Without extra dimensions the corresponding divergences are only logarithmic and hence independent of the loop parameterization.}. To do so, we demand gauge invariance, with regard to the Yang-Mills gauge group, of the one-loop counterterms and require that all bubbles, triangles, etc.\ are parameterized in a similar manner. This completely fixes the choice of the parameterization of the loops.

We illustrate our approach by presenting the explicit parameterization dependence of the gravitational contribution to the fermionic two-point function. The effective degree of divergence of the bubble graph is $3+\de$, in contrast to all further diagrams whose degree of divergence is at most $2+\de$. Consequently, also the regularized value of the leading divergence (of degree $2+\de$) depends on the chosen loop parameterization. Using a general distribution of the external momentum $q_\mu$ over the two arms of the bubble, the leading divergent term of the one-loop contributions to the two-point function is
\begin{equation}\label{eq:lead_ferm2pt}
\raisebox{-.9cm}{\includegraphics{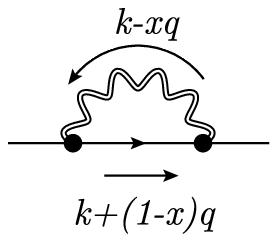}} +\raisebox{-.02cm}{\includegraphics{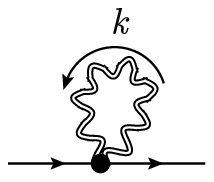}}\quad\raisebox{+0.3cm}{$\left.\parbox[m][2cm]{0cm}{}\right\vert_{\text{~leading order}}$}= \ka^2 \s{q} \frac{3}{4}\left[ \frac{11+9\de}{\de+2} +x\frac{4-\de}{2(\de+4)}\right]\sum_{\vec{n}}\pint{k}\frac{1}{k^2-m_{\vec{n}}^2}\,,
\end{equation}
with  $x\in \mathbb{R}$
parameterizing the fraction of the external momentum flowing on the graviton line. The requirement of gauge invariance of the counterterms now determines the value of $x$.

\begin{figure}[t]
\includegraphics{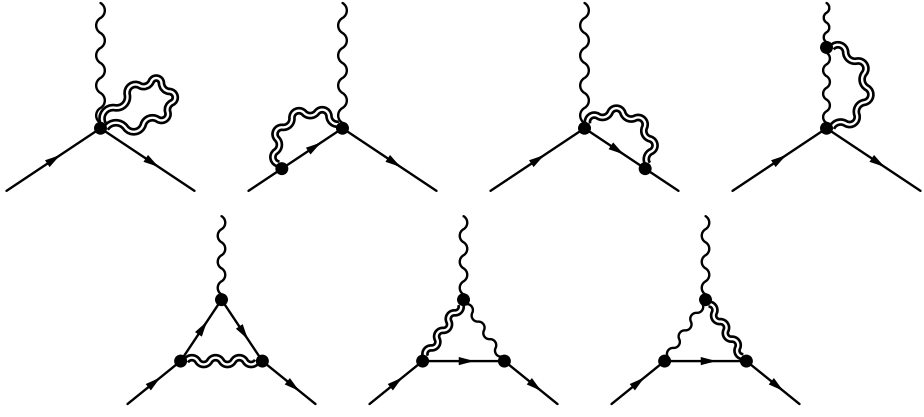}
\caption{One-loop diagrams for the proper two-fermion--one-gauge-boson vertex.}\label{fig:ferm_3pt}
\end{figure}

This leading term will contribute to the field-strength renormalization of the fermion $Z_{\psi}$. 
As discussed in \cite{ebert2008gcr} and \cite{schuster2008dpl}, due to the absence of a coupling between Yang-Mills ghosts and gravitons the Slavnov-Taylor identities require the field-strength and the gauge field vertex renormalization of a given field to be identical at order $\ka^2$:
\begin{equation}
 Z_{\psi}\Bigr|_{\O(\ka^2)} = Z_{\overline{\psi}A \psi}\Bigr|_{\O(\ka^2)}\,.
\end{equation}
 The effective degree of divergence of the gravitational one-loop diagrams contributing to the fermion--gluon vertex in figure~\ref{fig:ferm_3pt} is at most $2+\de$ and thus their contribution to $Z_{\overline{\psi}A\psi}$ is independent of the momentum parameterization:
\begin{align}\label{eq:lead_ferm3pt}
 \raisebox{-1cm}{\includegraphics{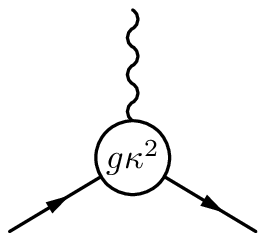}}\raisebox{+0.1cm}{$\left.\parbox[m][2.5cm]{0cm}{}\right\vert_{\text{~leading order}}$}
=g \ka^2  \ga^\mu \frac{3(11+9\de)}{4(\de+2)} \sum_{\vec{n}}\pint{k}\frac{1}{k^2-m_{\vec{n}}^2} \,.\\*[-1ex]\notag
\end{align}
 Comparison of this result with \eqref{eq:lead_ferm2pt} and the requirement of the Slavnov-Taylor identities to be satisfied now fixes $x=0$. Once a loop-momentum parameterization is chosen all power-like divergences are fixed. As the regularized integrals will depend on the chosen momentum parameterization, this particular gauge compatible parameterization has to be part of our regularization scheme. For all graphs with a bubble loop we applied a parameterization with no external momentum flowing on the graviton line.

Regularizing the remaining integrals and sums, one has to bear in mind that the Kaluza-Klein masses are the discrete values of the extra-dimensional components of the graviton momentum. Hence, sum and integral have to be regularized together and not separately. In order to achieve this, it is convenient to replace the sum over the Kaluza-Klein modes by a $\de$-dimensional integral\footnote{The difference is suppressed by $(R\La)^{-1}$ and hence negligible in our calculation.}. The resulting $4+\de$ dimensional momentum integral is treated in the Wilsonian sense and we integrate out a Euclidean momentum shell from the low-energy reference scale $\mu$ to the UV-cut-off scale $\La$. 

To determine the leading and next-to-leading divergences, investigated here, only the following two integrals are required:
\begin{align}
 		 \sum_{\vec{n}}\pint{k}\frac{1}{k^2-m_{\vec{n}}^2} & \xrightarrow{\text{Regularization}}
				-\frac{i\pi^{\de/2}R^\de}{8\pi^2\Gamma\left(\frac{\de}{2}+2\right)}\frac{\La^{\de+2}-\mu^{\de+2}}{\de+2}\label{eq:int1}\\
 \quad \sum_{\vec{n}}\pint{k}\frac{1}{k^2(k^2-m_{\vec{n}}^2)} & \xrightarrow{\text{Regularization}} \frac{i\pi^{\de/2}R^\de}{8\pi^2\Gamma\left(\frac{\de}{2}+1\right)}\frac{\La^{\de}-\mu^{\de}}{\de}\,.\label{eq:int2}
\end{align}

Using the fixed loop parameterization and the regularized expressions \eqref{eq:int1}, \eqref{eq:int2} one finds
\begin{equation}\label{eq:ferm_2pt}
\raisebox{-.4cm}{\includegraphics[height=1.5cm]{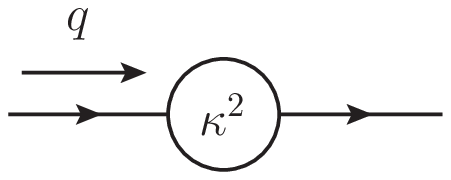}}
 =\frac{i}{(4\pi)^{1+\de/2}\Gamma\left(\frac{\de}{2}+2\right)}\Biggl[ -\frac{3(11+9\de)}{(\de+2)^2}\s{q}\frac{\Lambda^{\de+2}-\mu^{\de+2}}{\emS[\delta+2]} 
 - \frac{16-\de}{8 \de}\s{q}\s{q}\s{q}\frac{\Lambda^{\de}-\mu^{\de}}{\emS[\delta+2]} \Biggr] + \cdots
\end{equation}
for the proper two-point function. It is important to notice that all the higher derivative terms \eqref{eq:ferm_op} give the same contribution to the fermion two-point function. Hence, \eqref{eq:ferm_2pt} only fixes the sum of their coefficients. In order to specify the higher derivative counterterms we additionally have to investigate the vertex corrections.

To determine the subleading part of the one-loop vertex diagrams in figure~\ref{fig:ferm_3pt}, the parameterization of the triangle also diagrams has to be fixed. The universality of the regularization scheme -- one rule for all triangle graphs -- and the demand that the result can be expressed as a linear combination of the gauge-in\-vari\-ant operators \eqref{eq:ferm_op} fixes one unique parameterization of the loop momentum. Similarly to the bubbles, no external momentum is allowed on the graviton propagator. 
Applying the described regularization scheme, the gravitational one-loop contributions to the fermion--gluon vertex are
\begin{align}\label{eq:ferm_3pt}
\raisebox{-.8cm}[1.4cm][0cm]{\includegraphics[height=2.5cm]{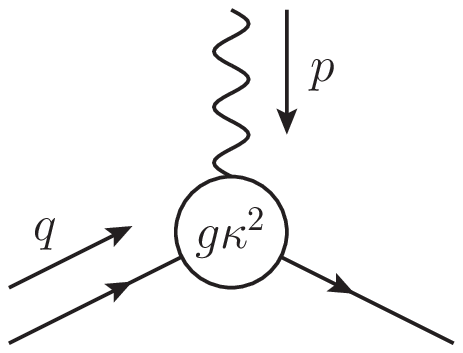}}
&=\frac{i\,g t^a}{(4\pi)^{1+\de/2}\Gamma\left(\frac{\de}{2}+2\right)}\Biggl[{}-\frac{3(11+9\de)}{(\de+2)^2}\gamma^{\mu}\frac{\Lambda^{\de+2}-\mu^{\de+2}}{\emS[\delta+2]} + \notag \\
&\=\phantom{\frac{i\,g t^a}{(4\pi)^{1+\de/2}\Gamma\left(\frac{\de}{2}+2\right)}\Biggl[}{}+\biggl\{{}-\frac{16-\de}{4}\s{q}q^{\mu}+\frac{32+25\de}{4}\s{q}p^\mu-\frac{80+49\de}{8}\s{q}\s{p}\gamma^\mu - \\*
&\=\phantom{\frac{i\,gt^a}{(4\pi)^{1+\de/2}\Gamma\left(\frac{\de}{2}+2\right)}\Biggl[{}+\biggl\{}{}-6(2+\de)\s{p}q^\mu-\frac{17(2+\de)}{6}\s{p}p^\mu-\frac{16-\de}{8}\gamma^\mu q^2+\notag\\
&\=\phantom{\frac{i\,gt^a}{(4\pi)^{1+\de/2}\Gamma\left(\frac{\de}{2}+2\right)}\Biggl[{}+\biggl\{}{}+\frac{32+25\de}{4}\gamma^\mu qp
		+\frac{88+71\de}{24}\gamma^\mu p^2\biggr\}\frac{\Lambda^{\de}-\mu^{\de}}{\delta\emS[\delta+2]}\Biggr]+\cdots\; \notag
\end{align}
and determine together with \eqref{eq:ferm_2pt} the higher derivative counterterms. The gravitational two-point \eqref{eq:ferm_2pt} and vertex \eqref{eq:ferm_3pt} diagrams, correspond to the counterterms
\begin{multline}\label{eq:ferm_ct}
 \L^\text{c.t.}_f=\frac{i}{(4\pi)^{1+\de/2}\Gamma\left(\frac{\de}{2}+2\right)}\,\overline{\psi}\Biggl[\frac{3(11+9\de)}{(\de+2)^2}\frac{\Lambda^{\de+2}-\mu^{\de+2}}{\emS[\delta+2]}\s{D} 
		+ \\
		+\Bigl\{-(10+\tfrac{49}{8}\de)\s{D}\s{D}\s{D}-(\tfrac{58}{3}+\tfrac{143}{12}\de)\s{D}D^2+(\tfrac{41}{3}+\tfrac{109}{12}\de)D^2\s{D}
		+(\tfrac{41}{3}
		+\tfrac{109}{12}\de)D_\mu \s{D}D^\mu\Bigr\}\frac{\Lambda^{\de}-\mu^{\de}}{\de\,\emS[\delta+2]}\Biggr]\psi
\end{multline}
which have to be added to the Lagrangian in order to renormalize the theory at one-loop order. In contrast to the results of \cite{wu2008tsl} we find that all higher derivative dimension-six operators are renormalized by gravitational one-loop corrections. An interpretation as fermionic Lee-Wick theory is therefore not possible due to the appearance of unitarity violating operators. For $4+0$ dimensions this result was successfully checked using dimensional regularization \cite{schuster2008dpl}. Note that for a nonvanishing fermion mass $m_{\scriptscriptstyle\psi}$ there will be an additional dimension-five term which also spoils the connection to the Lee-Wick standard model.

\section{Scalars}
The higher derivative counterterms for the charged scalar multiplet is obtained in the same manner as for the fermion. We start with the Lagrangian of the minimally coupled scalar field:
\begin{equation}
 \L_s=\sqrt{-g}g^{\mu\nu}(D_\mu\phi)^\dagger D_\nu \phi\qquad\qquad \text{with}\qquad\qquad D_\mu=\del_\mu-igA_\mu
\end{equation}
 and extract the Feynman rules for its coupling to the graviton from the expansion of the Lagrangian in orders of the gravitational coupling $\kappa$
\begin{equation}
 \L_s = (D_\mu\phi)^\dagger D^\mu\phi +\kappa\left[\tfrac{1}{2}\eta^{\mu\nu}h-h^{\mu\nu}\right](D_\mu\phi)^\dagger D_\nu\phi 
	+\kappa^2\bigl[\tfrac{1}{8}(h^2-2h^{\alpha\beta}h_{\alpha\beta})\eta^{\mu\nu}+h^{\mu\alpha}h^\nu_\alpha-\tfrac{1}{2}hh^{\mu\nu}\bigr](D_\mu\phi)^\dagger D_\nu\phi+
	\O(\kappa^3)\,.
\end{equation}
Explicit formulas can be found in \cite{schuster2008dpl}. In the case of the scalar field only three higher derivative operators are linearly independent. We choose the following basis for the counterterms:
\begin{equation}
(D^2\phi)^\dagger D^2\phi\,, \qquad ig(D_\mu\phi)^\dagger F^{\mu\nu} D_\nu \phi\,, \quad\text{and}\quad g^2\phi^\dagger F^{\mu\nu}F_{\mu\nu} \phi \,.
\end{equation}
Note that every term in the second operator contains at least one gauge field and every term in the third operator contains at least two gauge fields, thus beside the two-point function the two-scalar--one-gauge-boson and two-scalar--two-gauge-boson amplitudes are needed to completely determine the higher derivative counterterms. 

\begin{figure}
\includegraphics{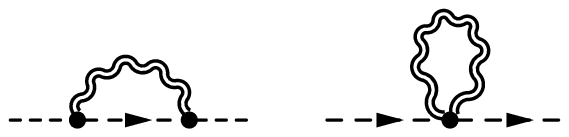}\vspace{-1cm}
\caption{One-loop diagrams for the proper scalar two-point function.}\label{fig:scalar_2pt}
\end{figure}
We start with the self-energy whose gravitational part is determined by the diagrams in figure~\ref{fig:scalar_2pt}. The UV-divergences are regularized in the same manner as before and the calculations show that again only the loop-parameterization without any external momentum flowing on the graviton line yield gauge-invariant results:
\begin{equation}\label{eq:scalar_2pt}
\raisebox{-.4cm}{\includegraphics[height=1.5cm]{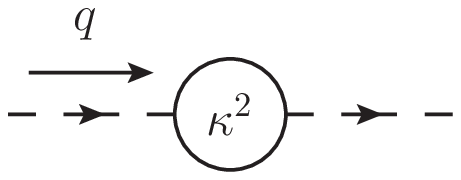}}
=\frac{i}{(4\pi)^{1+\de/2}\Gamma\left(\frac{\de}{2}+2\right)}\Biggr[-\frac{2(8+5\de)}{(\de+2)^2}q^2\frac{\Lambda^{\de+2}-\mu^{\de+2}}{\emS[\delta+2]} 
 + q^4\frac{\Lambda^{\de}-\mu^{\de}}{\emS[\delta+2]} \Biggr]+\cdots \,.
\end{equation}
Note that the higher derivative divergence $\sim q^4$ vanishes if extra-dimensions are absent, hence in four dimensions there is no Lee-Wick counterterm $(D^2\phi)^\dagger D^2\phi$. 

\begin{figure}
\includegraphics{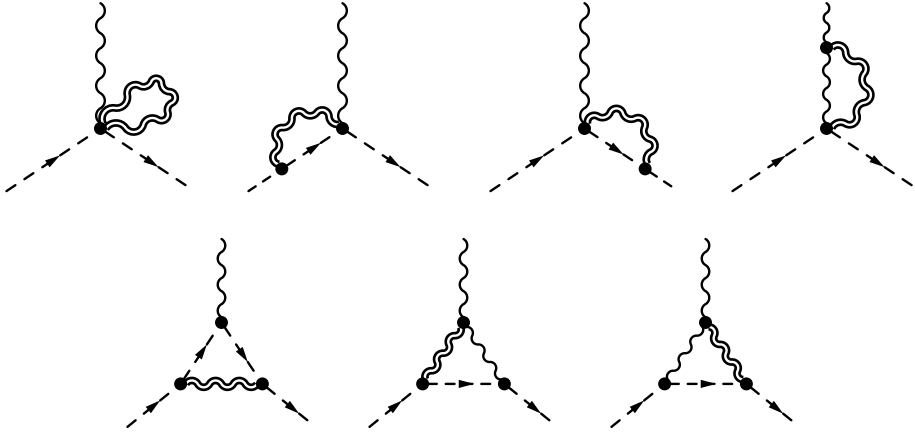}
\caption{One-loop diagrams for the proper two-scalar--one-gauge-field vertex.}\label{fig:scalar_3pt}
\end{figure}
The one-loop diagrams for the two-scalar--one-gauge-field vertex are listed in figure~\ref{fig:scalar_3pt}. Their sum is
\begin{equation}\label{eq:scalar_3pt}
\begin{aligned}
\raisebox{-.8cm}[1.4cm][0cm]{\includegraphics[height=2.5cm]{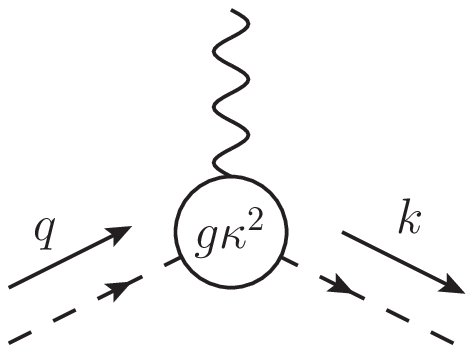}}
=\frac{i\,gt^a}{(4\pi)^{1+\de/2}\Gamma\left(\frac{\de}{2}+2\right)}\Biggl[&{}-\frac{2(8+5\de)}{(\de+2)^2}(q+k)^\mu \frac{\Lambda^{\de+2}-\mu^{\de+2}}{\emS[\delta+2]} + \\
		&+\biggl\{{} (q^2 +\tfrac{1}{3}qk +\tfrac{2}{3}k^2) q^\mu 
			+ (k^2 +\tfrac{1}{3}kq +\tfrac{2}{3}q^2) k^\mu \biggr\}\frac{\Lambda^{\de}-\mu^{\de}}{\emS[\delta+2]}\Biggr]+\cdots .
\end{aligned}
\end{equation}
Again all dimension-six contributions vanish for $\delta=0$ and hence in four dimensions there is also no counterterm of the form $ig(D_\mu\phi)^\dagger F^{\mu\nu} D_\nu \phi$.

\begin{figure}
\includegraphics{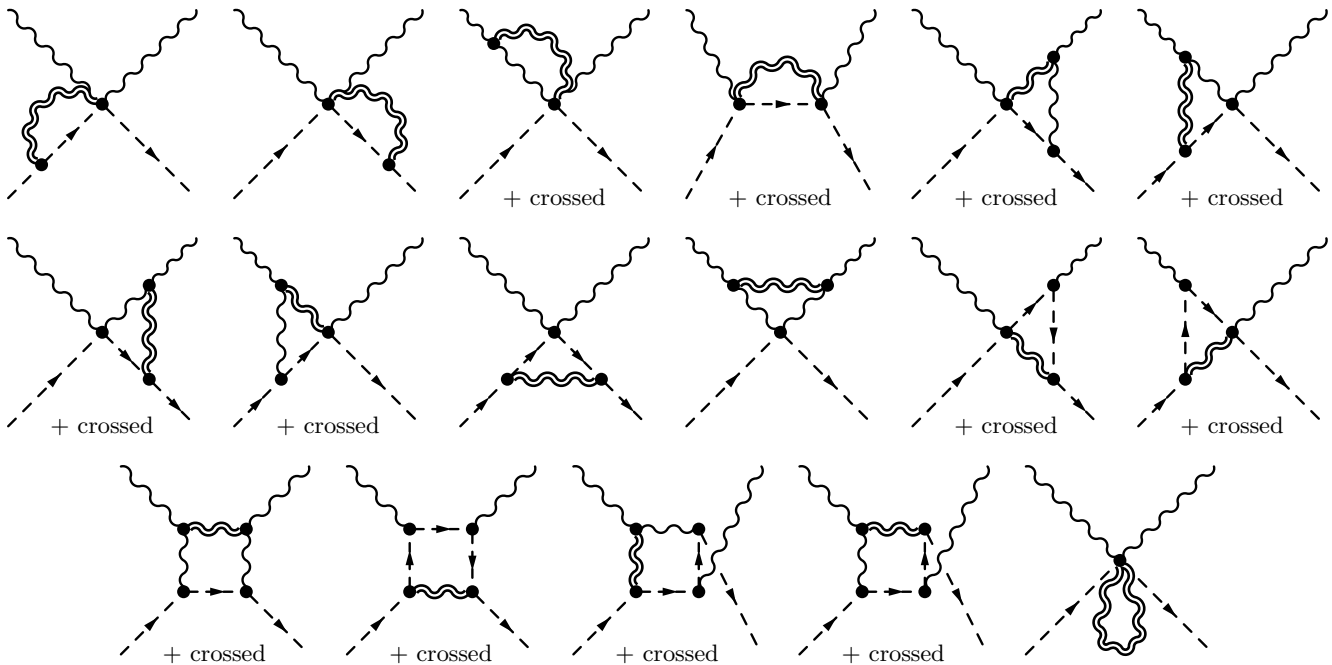}
\caption{One-loop diagrams for the two-scalar--two-photon vertex. Here ``+~crossed'' refers to the corresponding diagram with exchanged external photon lines.}\label{fig:scalar_4pt}
\end{figure}

Finally, to determine the counterterm for the operator $\phi^\dagger F^{\mu\nu}F_{\mu\nu} \phi$ the two-scalar--two-gauge-field amplitude is necessary. To simplify the calculation, it is done for an Abelian theory with the Abelian coupling constant $e$ which already leads to the 29 one-loop  diagrams listed in figure~\ref{fig:scalar_4pt}. All these diagrams sum up to
\begin{equation}\label{eq:scalar_4pt}
\begin{aligned}
\raisebox{-1cm}[1.8cm][0.5cm]{\includegraphics[height=3cm]{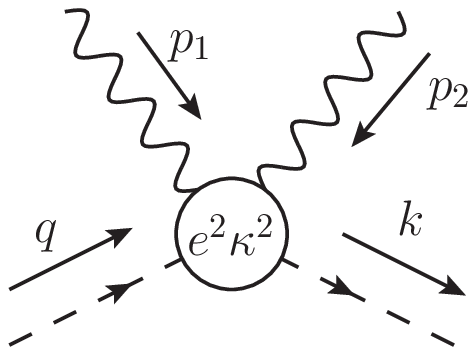}}
=\frac{i\,e^2}{(4\pi)^{1+\de/2}\Gamma\left(\frac{\de}{2}+2\right)}\Biggl[&{}-\frac{4(8+5\de)}{(\de+2)^2}\eta^{\mu \nu} \frac{\Lambda^{\de+2}-\mu^{\de+2}}{\emS[\delta+2]} + \\
	&+\biggl\{{} 2 (q+k)^\mu (q+k)^\nu + (q+k)^2 \eta^{\mu \nu} + \\
	&\phantom{+\biggl\{}+\tfrac{1}{3} p_1^\mu p_1^\nu -2 p_1^\nu p_2^\mu +\tfrac{1}{3} p_2^\mu p_2^\nu + \\
	&\phantom{+\biggl\{}+\eta^{\mu \nu}(\tfrac{2}{3} p_1^2 +2 p_1 p_2 +\tfrac{2}{3} p_2^2) \biggr\}\frac{\Lambda^{\de}-\mu^{\de}}{\emS[\delta+2]}\Biggr]+\cdots\, .
\end{aligned}
\end{equation}
Since gravity is insensitive to the particular gauge group chosen and because of the identical structure of the scalar field higher derivative operators for Abelian and non-Abelian groups, this result can easily be generalized to a non-Abelian theory. The nontrivial consistency of the thus obtained non-Abelian two-scalar--two-gauge-field vertex with the results \eqref{eq:scalar_2pt} and \eqref{eq:scalar_3pt} further justifies this simplification.

From \eqref{eq:scalar_2pt}--\eqref{eq:scalar_4pt} we obtain the gravitational counterterms involving two scalars\footnote{For the scalar field there are additional dimension-six operators, e.\,g. $(\phi^\dagger D_{\mu}\phi)^2$ or $(\phi^\dagger\phi)^3$, which are expected to be renormalized by gravity effects too.}
\begin{multline}\label{eq:scalar_ct}
  \L^\text{c.t.}_s=\frac{i}{(4\pi)^{1+\de/2}\Gamma\left(\frac{\de}{2}+2\right)}\biggl[ \frac{2(8+5\de)}{(\de+2)^2}(D_\mu\phi)^\dagger D^\mu\phi \frac{\Lambda^{\de+2}-\mu^{\de+2}}{\emS[\delta+2]}-\\
		-\left\{ (D^2\phi)^\dagger D^2\phi -\tfrac{1}{3}ig(D_\mu\phi)^\dagger F^{\mu\nu} D_\nu \phi +\tfrac{1}{6} g^2\phi^\dagger F^{\mu\nu}F_{\mu\nu} \phi \right\} 
			\frac{\Lambda^{\de}-\mu^{\de}}{\emS[\delta+2]} \biggr].
\end{multline}
Similarly to the fermions, for the scalar field in a general setup, no higher derivative terms can be excluded if quantum gravity effects are incorporated. Again this contradicts the possible connection between gravity and the Lee-Wick standard model claimed by \cite{wu2008tsl}. Interestingly, all the dimension-six counterterms only appear in the extra-di\-men\-sional case. The absence of these higher derivative counterterms in four dimensions has been checked using dimensional regularization \cite{schuster2008dpl}. 
\section{Conclusions}
We have established a cut-off regularization applicable in extra-dimensional scenarios which involves a cut-off of the $(4+\delta)$-dimensional\footnote{This regularization is also applicable in $D=(d+\delta)$ dimensions \cite{ebert2008gcr}.} momentum and a fixed parameterization of the loops, where the graviton propagator does not carry any external momenta. This particular parameterization is completely fixed by the demand of gauge invariance of the counterterms and by requiring that all bubbles, triangles, etc.\ are parameterized in identical manner.

Applying this regularization we determined the dimension-six one-loop counterterms involving two fermions or two scalars to be a linear combination of all possible operators \eqref{eq:ferm_ct}, \eqref{eq:scalar_ct} and not only the Lee-Wick terms, as in the gauge sector \eqref{eq:gauge_ct}. This contradicts the results of \cite{wu2008tsl} and leads to the conclusion that there is no connection between the gravitational one-loop counterterms of standard model matter and the Lee-Wick standard model. It is important to note that the appearance of the Lee-Wick term for gauge fields as the only gauge field counterterm, although it is tempting, does not allow to conclude the existence of the massive particle associated with the Lee-Wick term in the Lee-Wick standard model, as has been done by \cite{Wu:2007lw}. In the  context considered here, the Lee-Wick term is only one term in an infinite series of counterterms of a nonrenormalizable field theory. One might argue that the higher order terms are exactly such that they correspond to further new particles, as in the higher derivative Lee-Wick standard model recently proposed by Carone and Lebed \cite{carone:2008iw}. However, it is in general not possible to conclude the existence of further particles from the appearance of special counterterms in an effective field theory, because these terms are only the residual low-energy effects of the unknown physics at high energies.
\section*{Acknowledgments}
We thank A.~A.\  Slavnov, and D.\  Ebert for interesting discussions and are especially grateful to J.\  Plefka for useful comments and his encouragement.
This work was supported by the Volkswagen Foundation.\\
Our computation made use of the symbolic manipulation system \textsc{Form} \cite{Form}.

\end{document}